\pdfoutput=1 
\documentclass{article}

\usepackage[left=25mm,right=25mm,top=25mm,bottom=25mm,a4paper]{geometry}
 
\usepackage{microtype}\usepackage{textcomp}
\usepackage{times}

\usepackage[sort&compress,numbers]{natbib}	

\usepackage{amsthm}
\usepackage{amsmath}
\usepackage{amssymb}

\usepackage{xspace}

\usepackage{graphicx}
\usepackage[super]{nth}

\newcommand{\xth}{\ensuremath{^{\text{th}}}\xspace}

\title{A Note on the Period Enforcer Algorithm  \\ for Self-Suspending Tasks}

\date{}

\author{Jian-Jia Chen \\ TU Dortmund University, Germany \\ \texttt{jian-jia.chen@cs.uni-dortmund.de} \and Bj\"orn B.\ Brandenburg \\ Max Planck Institute for Software Systems (MPI-SWS) \\ \texttt{bbb@mpi-sws.org}}

\begin{document}

\maketitle
\begin{abstract}
The \emph{period enforcer} algorithm for self-suspending real-time tasks is a technique for suppressing the ``back-to-back'' scheduling penalty associated with deferred execution. Originally proposed in 1991, the algorithm has attracted renewed interest in recent years. This note revisits the algorithm in the light of recent developments in the analysis of self-suspending tasks, carefully re-examines and explains its underlying assumptions and limitations, and points out three observations that have not been made in the literature to date: \textbf{(i)}~period enforcement is not strictly superior (compared to the base case without enforcement) as it can cause deadline misses in self-suspending task sets that are schedulable without enforcement; \textbf{(ii)}~to match the assumptions underlying the analysis of the period enforcer, a schedulability analysis of self-suspending tasks subject to period enforcement requires a task set  transformation for which no solution is known  in the general case, and which is subject to exponential time complexity (with current techniques) in the limited case of a single self-suspending task; and \textbf{(iii)}~the period enforcer algorithm is incompatible with all existing analyses of suspension-based locking protocols, and can in fact cause ever-increasing suspension times until a deadline is missed.
\end{abstract}

\section{Introduction} 
When real-time tasks suspend themselves (due to blocking I/O, lock contention, \textit{etc.}), they defer a part of their execution to be processed at a later time. A consequence of such deferred execution is a potential interference penalty for lower-priority tasks~\cite{LSS:87,LSST:91,Ra:90,ABRTW:93,SLS:95,WC16-suspend-DATE,ecrts15nelissen}. This penalty, which is maximized when a task defers the completion of one job just until the release of the next job, can manifest as response-time increases and thus may lead to deadline misses.

To avoid such detrimental effects,  Rajkumar \cite{Raj:suspension1991} proposed the \emph{period enforcer} algorithm,  a technique to control (or shape) the processor demand of self-suspending tasks on uniprocessors and partitioned multiprocessors under preemptive fixed-priority scheduling. In a nutshell, the period enforcer algorithm artificially increases the length of certain suspensions whenever a task's activation pattern carries the risk of inducing undue interference in lower-priority tasks. 

The period enforcer algorithm is worth a second look for a number of reasons. First, in the words of Rajkumar, it ``forces tasks to behave like ideal periodic tasks from the scheduling point of view with no associated scheduling penalties''~\cite{Raj:suspension1991}, which is obviously highly desirable in many practical applications in which self-suspensions are inevitable (e.g., when offloading computations to co-processors such as GPUs or DSPs). Second, the later-proposed, but more widely-known \emph{release guard} algorithm~\cite{SL:96} uses a technique quite similar to period enforcement to control scheduling penalties due to release jitter in distributed systems. The period enforcer algorithm has also attracted renewed attention in recent years and has been discussed in several current works  (e.g.,~\cite{DBLP:conf/rtss/ChenL14,LNR:09,LR:10,Lak:11,LC:14,KANR:13,HY:11,CA:09,CA:10,CA:10b}), at times controversially~\cite{BA:08a}. And last but not least, the period enforcer algorithm plays a significant role in Rajkumar's seminal book on   real-time  synchronization~\cite{Raj:91}. 

In this note, we revisit the period enforcer \cite{Raj:suspension1991} to carefully re-examine and explain its underlying assumptions and limitations, and to point out potential misconceptions.  The main contributions are three observations that, to the best of our knowledge, have not been previously reported in the literature on real-time systems:
\begin{enumerate}
	\item period enforcement can be a cause of deadline misses in self-suspending task sets that are otherwise schedulable (Section~\ref{sec:unschedulable}); 
	\item to match the assumptions underlying the analysis of the period enforcer, a schedulability analysis of self-suspending tasks subject to period enforcement requires a task set  transformation for which no solution is known  in the general case, and which  is subject to exponential time complexity (with current techniques)  in the limited case of a single self-suspending task (Section~\ref{sec:convert}); and
	\item the period enforcer algorithm is incompatible with all existing analyses of suspension-based locking protocols, and can in fact cause ever-increasing suspension times until a deadline is missed (Section~\ref{sec:locking}).
\end{enumerate}

We introduce the needed background in Section~\ref{sec:prelim}, restate our contributions more precisely in Section~\ref{sec:questions}, and then establish the three above  observations in detail in Sections \ref{sec:unschedulable}--\ref{sec:locking} before concluding in Section \ref{sec:conclusion}.

\section{Preliminaries}
\label{sec:prelim}

The period enforcer algorithm~\cite{Raj:suspension1991} applies to self-suspending tasks on uniprocessors under fixed-priority scheduling, and hence by extension also to multiprocessors under partitioned fixed-priority scheduling (where tasks are statically assigned to processors and each processor is scheduled as a uniprocessor). In this section, we review the underlying task model (Section \ref{sec:taskmodel}), introduce the period enforcer algorithm (Section \ref{sec:pe}), summarize its analysis (Section \ref{sec:classic-analysis}), and finally restate our observations in more precise terms (Section \ref{sec:questions}).

\subsection{Task Models}
\label{sec:taskmodel}

Since the analysis of the period enforcer requires reasoning about different task models and their relationships, we carefully introduce and precisely define the relevant models in this section.

\subsubsection{Periodic Tasks}

The most basic and best understood task model is the \emph{periodic task model} due to Liu and Layland~\cite{LL:73}. In this model, each task $\tau_i$ is characterized as a tuple $(C_i,T_i)$, where $C_i$ denotes an upper bound on the total execution time of any job of $\tau_i$ and $T_i$ denotes the (exact) \emph{inter-arrival time} (or \emph{period}) of $\tau_i$. Each such periodic task $\tau_i$ releases a job at time~0, and periodically every $T_i$ units thereafter. Each job must finish by the time the next arrives. Importantly, Liu and Layland assume both that the $k$\xth job of $\tau_i$ arrives \emph{exactly} at time $(k-1)\times T_i$, and that an incomplete job is \emph{always} available for execution (i.e., jobs never block on I/O or locks).

A straightforward generalization of the periodic task model is to introduce an explicit \emph{relative deadline} parameter $D_i$. In this case, each task is represented by a three-tuple $(C_i, T_i, D_i)$, with the interpretation that every job of $\tau_i$ must finish within $D_i$ time units after its release. Task $\tau_i$ is said to have an \emph{implicit deadline}  if $D_i = T_i$, a \emph{constrained deadline} if $D_i \leq T_i$, and an \emph{arbitrary deadline} otherwise. We primarily consider implicit deadlines in this note.

\subsubsection{Sporadic Tasks}

Mok~\cite{Mo:83} introduced the \emph{sporadic task model}, a widely used generalization of the periodic task model in which each task $\tau_i$ is still specified by a tuple $(C_i, T_i, D_i)$. However, the sporadic task model relaxes the inter-arrival constraint $T_i$ to specify a \emph{minimum} (rather than an exact) separation between jobs. In this interpretation, the first job is not necessarily released at time~0, and the exact release times of future jobs cannot be predicted, which is an appropriate modeling assumption for event-triggered tasks.

On uniprocessors, the relaxation from periodic to sporadic job arrivals does not introduce additional pessimism:\footnote{Assuming that all periodic tasks synchronously release a job at time zero.} since any two jobs of a sporadic task $\tau_i$ are  known to be released \emph{at least} $T_i$ time units apart, the sporadic task model~\cite{Mo:83} still allows for schedulability analysis that is as accurate as Liu and Layland's analysis of periodic tasks~\cite{LL:73}. 

Mok retained the assumption that incomplete jobs are always ready for execution (i.e., no suspensions), and that jobs, once released, are \emph{immediately} available for execution.

\subsubsection{Release Jitter}
\label{sec:jitter}

The latter assumption --- immediate availability for execution --- is inappropriate in many practical systems (especially in networked systems) if events (e.g., messages) that trigger job releases can incur non-negligible delays (e.g., network congestion). Such delays in task activation can be accounted for by introducing a notion of \emph{release jitter}. To this end, each task is represented by a four-tuple $(C_i, J_i, T_i, D_i)$, where the parameter $J_i$ is a bound on the maximum time that a job remains unavailable for execution after it should have started to run. Release jitter can be incorporated in both the periodic and the sporadic task models.

In the presence of release jitter, the terms ``job arrival'' and ``job release,'' which are often used interchangeably, take on distinct meanings: a job's \emph{arrival time} denotes the point in time when it actually becomes available for execution, whereas a job's \emph{release time} is the instant that is relevant for the (minimum) inter-arrival time constraint. Any job of task $\tau_i$  \emph{arrives} at most $J_i$ time units after it is \emph{released}.

Notably, non-zero release jitter \emph{does} cause additional pessimism: in the worst case, two consecutive jobs of a task $\tau_i$ can be separated by as little as $T_i - J_i$ time units (if the earlier job incurs maximum release jitter and the successor job incurs none). As a result, a task may ``carry in'' some additional work into a given interval. Taking this effect into account, Audsley et al.~\cite{ABRTW:93} developed a response-time analysis for sporadic and periodic constrained-deadline tasks subject to release jitter under preemptive fixed-priority scheduling.

However, even in the presence of release jitter, a key assumption remains that jobs do not self-suspend (e.g., wait for I/O).\footnote{Audsley et al.~\cite{ABRTW:93} do present a response-time analysis that takes into account a limited form of suspensions due to semaphores (``blocking''). However, their analysis does not apply to general self-suspensions (i.e., the kind of self-suspensions targeted by the period enforcer algorithm)  and is not relevant in the context of this paper.} That is, Audsley et al.~\cite{ABRTW:93} assume that, once a job has arrived, it continuously remains available for dispatching until it completes. This restriction is removed next.

\subsubsection{Self-Suspending Tasks}
\label{sec:dynamic}

When a job \emph{self-suspends}, it becomes unavailable for execution until some external event occurs (e.g., a disk I/O operation completes, a network packet arrives, a co-processor signals completion, \textit{etc.}). This has the effect of \emph{deferring} (a part of) the job's processing requirement until the time that it \emph{resumes} from its suspension, which causes massive analytical difficulties~\cite{LSS:87,LSST:91,Ra:90,ABRTW:93,SLS:95,WC16-suspend-DATE,ecrts15nelissen,Ri:04,Raj:suspension1991,Chen2016}. 

To date, the real-time literature on self-suspensions has focused on two task models: the \emph{dynamic} self-suspension model, which we discuss first,  and the  (\emph{multi}-)\emph{segmented} suspension model, which we discuss next in Section~\ref{sec:segmented}.  Self-suspensions can arise in both periodic and sporadic tasks (i.e., both interpretations of the $T_i$ parameter are possible). The observations that we make in this note apply equally to both periodic and sporadic tasks; for convenience, we focus primarily on periodic tasks.

The dynamic self-suspending task model characterizes each
task $\tau_i$ as a four-tuple $(C_i,S_i,T_i,D_i)$: the parameters $C_i$, $T_i$, and $D_i$ have their usual meaning (i.e., as in the periodic and sporadic task models), and $S_i$ denotes an upper bound on the total self-suspension time of any job of $\tau_i$. The dynamic self-suspension model does not impose a bound on the maximum number of self-suspensions, nor does it make any assumptions as to where during a job's execution self-suspensions occur. That is, how often a job defers its execution, when it does so, and how much of its execution it defers may vary unpredictably from job to job.

Allowing tasks to self-suspend can impose substantial scheduling penalties (an example is provided shortly in Section~\ref{sec:pe}) and greatly complicates  schedulability analysis (e.g., see~\cite{ecrts15nelissen,Ri:04,Chen2016}). In particular, release jitter and self-suspensions are not interchangeable concepts and it is not safe~\cite{Chen2016,ecrts15nelissen} to simply substitute $J_i$ with $S_i$ in Audsley et al.'s analysis~\cite{ABRTW:93}. (Nonetheless, under the dynamic suspension model, it is possible for jobs of self-suspending tasks to defer their entire execution requirement, so self-suspensions can be seen as a generalization of release jitter.)

The period enforcer algorithm aims to mitigate the negative effects of self-suspensions. However, for reasons that will be explained in Section~\ref{sec:incompat}, the period enforcer algorithm cannot be meaningfully combined with the dynamic suspension model. Instead, it requires the segmented suspension model, which we discuss next.

\subsubsection{Segmented Self-Suspending Tasks}
\label{sec:segmented}

The (multi-)segmented self-suspending sporadic task model extends the  four-tuple $(C_i,S_i,T_i,D_i)$ by characterizing each self-suspending task as a fixed, finite linear sequence of computation and suspension intervals. These intervals are represented as a tuple
$(S_{i}^0,C_{i}^1,S_{i}^1,C_{i}^2,S_{i}^2,...,S_{i}^{m_i-1},C_{i}^{m_i})$, which is composed of $m_i$ computation segments separated by $m_i$ suspension intervals.

The first self-suspension segment $S_i^0$, prior to the first execution segment, is equivalent to release jitter (i.e., the parameter $J_i$ in Section~\ref{sec:jitter}). However, in much of the literature on the segmented self-suspending task model, the segment $S_i^0$ is assumed to be absent (i.e., $S_i^0 = 0$), such that there are only $m_i - 1$ suspension intervals (and jobs arrive jitter-free). Unless noted otherwise we adopt this convention.

We say that a segment \emph{arrives} when it becomes available for execution. The first computation segment arrives immediately when the job is released (unless $S_i^0 \neq 0$); the second computation segment (if any) arrives when the job resumes from its first self-suspension, \textit{etc.}

The advantage of the dynamic model (Section~\ref{sec:dynamic}) is that it is more flexible since it does not impose any assumptions on a task's control flow. The advantage of the segmented model is that it allows for more accurate analysis. The period enforcer algorithm and its analysis~\cite{Raj:suspension1991} applies (only) to the segmented model, as explained in Sections~\ref{sec:incompat} and~\ref{sec:classic-analysis}. 

A note on terminology: for the sake of consistency with the recent literature on self-suspensions in real-time systems, we favor the term ``segmented self-suspending tasks'' to refer to tasks under the just-introduced model. However,  Rajkumar's original description of the period enforcer~\cite{Raj:suspension1991} refers to such tasks as  \emph{deferrable tasks}, as it predates the widespread adoption of the former term. We use both terms interchangeably in this paper.

\subsubsection{Single-Segment Self-Suspending (aka Deferrable) Tasks}
\label{sec:single-segmented}
An important special case is segmented self-suspending tasks with exactly one self-suspension interval followed by exactly one computation segment ($m_i = 1$, $S_i^0 \neq 0$), which we refer to as \emph{single-segment self-suspending tasks}.  This special case is central to Rajkumar's original analysis of the period enforcer~\cite{Raj:suspension1991},  as we will explain in Section~\ref{sec:classic-analysis}.
Regarding terminology, Rajkumar~\cite{Raj:suspension1991} does not use a special term for single-segment self-suspending tasks, simply referring to them as deferrable tasks. To avoid ambiguity, we instead explicitly mention the ``single-segment'' qualifier.  

Note also that single-segment self-suspending sporadic tasks, which are  ``suspended'' only prior to commencing execution, are analytically fully equivalent to sporadic tasks subject to release jitter (i.e., the model described in Section~\ref{sec:jitter}). We nonetheless use the term ``single-segment self-suspending task,'' or interchangeably ``single-segment deferrable task,'' to remain close to Rajkumar's original description~\cite{Raj:suspension1991}, and to highlight the connection to the (multi-)segmented self-suspending task model (Section~\ref{sec:segmented}).

This concludes our review of relevant task models. Before reviewing the period enforcer and its original analysis, we briefly introduce some essential concepts.

\subsubsection{Assumptions, Busy Periods, and Task Set Transformations}
\label{sec:misc-defs}

We focus exclusively on preemptive fixed-priority scheduling in this note, as the period enforcer is explicitly designed for this setting. For simplicity, we assume that tasks are indexed in order of decreasing priority (i.e., $\tau_1$ is the highest-priority task). 

A key concept in the period enforcer's runtime rules (discussed next) is the notion of a \emph{level-$i$ busy interval}, which is a maximal interval during which  the processor executes only segments of tasks with priority $i$ or higher.

\newcommand{\tset}[0]{\mathcal{T}}

Finally, Rajkumar's original analysis~\cite{Raj:suspension1991} of the period enforcer is rooted in the concept of a \emph{task set transformation}. In general, such a task set transformation is simply a function $f$ that maps a given task set $\tset$ to a transformed task set $\tset' = f(\tset)$ \emph{such that $\tset'$ is schedulable \textbf{only if} the original task set $\tset$ is schedulable}, too. The basic idea is that such a transformation allows schedulability analysis by reduction: given a suitable transformation $f$, $\tset$ can be  \emph{indirectly} shown to be schedulable by computing $\tset' = f(\tset)$ and establishing that $\tset'$ is schedulable.

Importantly, the tasks in $\tset$ and $\tset'$ do \emph{not} have to be of the same task model, nor does the number of tasks have to remain the same (i.e.,  $|\tset| \neq |\tset'|$ is possible). Specifically, the task set transformation underlying the analysis of the period enforcer maps each \emph{multi-}segmented self-suspending task $\tau_i \in \tset$  to $m_i$ \emph{single-}segmented self-suspending tasks in $\tset'$ (i.e., $|\tset'| = \sum_{\tset} m_i$).

With these definitions in place, we can now introduce the period enforcer.

\subsection{The Period Enforcer Algorithm}
\label{sec:pe}

The period enforcer consists of two parts: a runtime rule that governs when each segment of a self-suspending task may be scheduled, and an (offline) analysis that may be used to assess the temporal correctness of a set of self-suspending tasks (Section~\ref{sec:segmented}) subject to period enforcement. Initially, we focus on the runtime rule (i.e., the actual period enforcer algorithm) and then review the corresponding original analysis thereafter in Section~\ref{sec:classic-analysis}.  We begin with a simple example that highlights the effect that the period enforcer is designed to control.

\subsubsection{The Problem: Back-to-Back Execution}

\begin{figure}[t]
  \centering
  \includegraphics[scale=1]{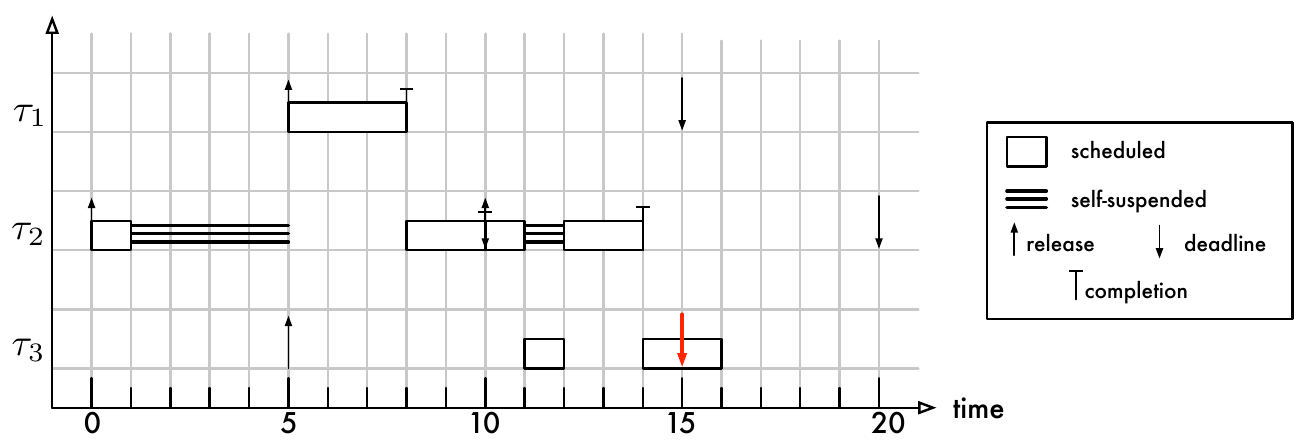}
  \caption{Example uniprocessor schedule (\emph{without} period enforcement) of three tasks $\tau_1$, $\tau_2$, and $\tau_3$ with periods $T_1 = T_2 = T_3 = 10$. Tasks $\tau_1$ and $\tau_3$ consist of a single computation segment ($C_1^1 = C_3^1 = 3$); task $\tau_2$ consists of two computation and one suspension segment ($C_2^1 = 1$, $S_2^1 = 4$, $C_2^2 = 2$). Jobs of tasks $\tau_1$ and $\tau_3$ are released just as $\tau_2$ resumes from its self-suspension at time~$5$. Without period enforcement, task $\tau_3$ misses a deadline at time~$15$ because the second job of  task $\tau_2$ suspends only briefly (for one time unit rather than four).}
  \label{fig:not-ok-without-period-enforcement}
\end{figure}

The scheduling penalty associated with self-suspensions is maximized when a task defers the completion of one job just until the release of the next job.  This effect is illustrated in Figure \ref{fig:not-ok-without-period-enforcement}, which shows a case in which the self-suspension of the higher-priority task $\tau_2$ from time~1 until time~5  results in a deadline miss of the lower-priority task $\tau_3$ at time~15.

The root cause is increased interference due to the ``back-to-back'' execution effect~\cite{LSS:87,LSST:91,Ra:90,ABRTW:93,SLS:95}. In the example shown in Figure \ref{fig:not-ok-without-period-enforcement}, two jobs of $\tau_2$ execute in close succession (i.e., separated by less than a period) because the second job, released at time~10, self-suspended for a (much) shorter duration than the first job.  Consequently,  $\tau_3$ suffers from increased interference when $\tau_2$'s second job resumes ``too soon'' at time~12 after having been suspended for only one time unit, rather than four time units like the first job of $\tau_2$. 

\subsubsection{The Period Enforcement Rule}

The key idea underlying the period enforcer algorithm is to artificially delay the execution of computation segments if a job resumes ``too soon.'' To this end,  the period enforcer determines for each computation segment an \emph{eligibility time}. If a segment resumes  before its eligibility time, the execution of the segment is delayed until the eligibility time is reached.

A segment's eligibility time is determined according to the following rule. Let $ET_{i,j}^k$ denote the eligibility time of the $k$\xth computation segment of the $j$\xth job of task $\tau_i$. Further, let $a^k_{i,j}$ denote the segment's arrival time. Finally, let $\mathit{busy}(\tau_i, t')$ denote the last time that a level-$i$ busy interval begins on or prior to time $t'$ (i.e., the processor executes only $\tau_i$ or higher-priority tasks throughout the interval $[\mathit{busy}(\tau_i, t'), t']$). The period enforcer algorithm defines the segment eligibility time of the $k$\xth segment as
\begin{align}\label{eq:ET-def}
	ET_{i,j}^k & = \max\left(ET_{i,j-1}^k + T_i,\ \mathit{busy}(\tau_i, a^k_{i,j})\right),
\end{align}
where $ET_{i,0}^k = -T_i$~\cite[Section 3.1]{Raj:suspension1991}. This simple and elegant rule has the desirable effect of avoiding all back-to-back execution, which can be easily observed with an example.

\subsubsection{Example: Avoiding Back-to-Back Execution}

\begin{figure}[t]
  \centering
  \includegraphics[scale=1]{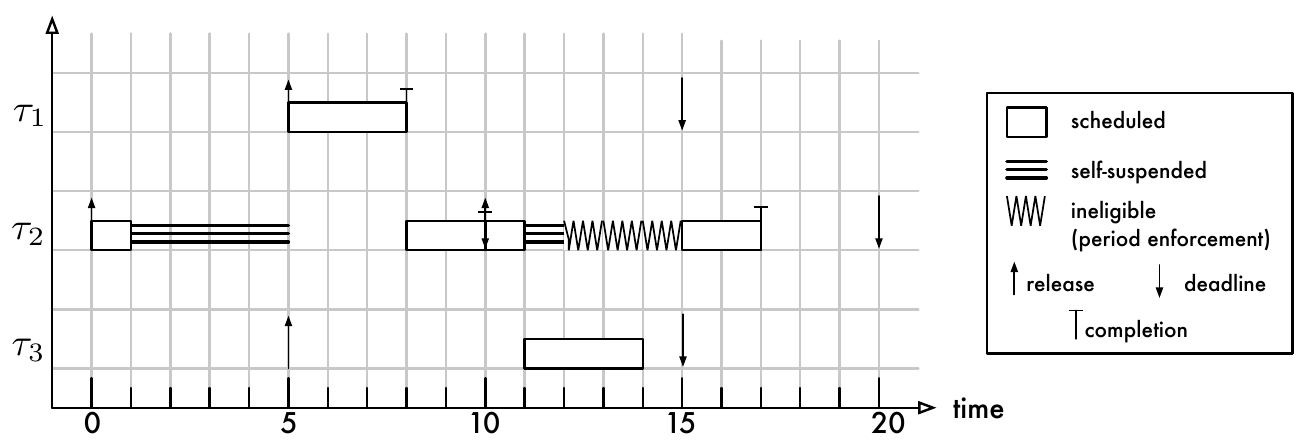}
  \caption{Example uniprocessor schedule \emph{with} period enforcement assuming the same scenario as depicted in Figure~\ref{fig:not-ok-without-period-enforcement}. With period enforcement, task $\tau_3$ does not miss a deadline because task $\tau_2$'s second computation segment is delayed until time~15 when it no longer imposes undue interference (i.e., it is prevented from resuming ``too soon'' at time~12).}
  \label{fig:ok-with-period-enforcement}
\end{figure}

Figure \ref{fig:ok-with-period-enforcement} illustrates how the definition of eligibility time in Equation~\eqref{eq:ET-def} restores the schedulability of the task set depicted in Figure~\ref{fig:not-ok-without-period-enforcement}. Consider the eligibility times of the second segment of task $\tau_2$.

By definition,  $ET_{2,0}^2 = -T_2 = -10$. At time~5, when the second computation segment of the first job resumes ($a_{2,1}^2 = 5$), we thus have
\begin{align*}
	ET_{2,1}^2 & = \max\left(-T_2 + T_2,\ \mathit{busy}(\tau_2, a_{2,1}^2\right) ) = \max(0, 5) = 5
\end{align*}
since the arrival of $\tau_2$'s second segment (and the release of $\tau_1$) starts a new level-2 busy interval at time $a_{2,1}^2 = 5$. The second segment of $\tau_2$'s first job is hence immediately eligible to execute; however, due to the presence of a pending higher-priority job, $\tau_2$ is not actually scheduled until time~8 (just as without period enforcement as depicted in Figure \ref{fig:not-ok-without-period-enforcement}).

The second segment of the second job of $\tau_2$ arrives at time $a_{2,2}^2 = 12$. In this case, the segment is \emph{not} immediately eligible to execute since
\begin{align*}
	ET_{2,2}^2 & = \max\left(ET_{2,1}^2 + T_2,\ \mathit{busy}(\tau_2, a_{2,2}^2\right) ) = \max(5 + 10, 12) = 15.
\end{align*}
Hence, the execution of $\tau_2$'s second computation segment does not start until time $ET_{2,2}^2 = 15$, which gives $\tau_3$ sufficient time to finish before its deadline at time~$15$.

The examples in Figures \ref{fig:not-ok-without-period-enforcement} and \ref{fig:ok-with-period-enforcement} suggest an intuition for the benefits provided by period enforcement: computation segments of a self-suspending task $\tau_i$ are forced to execute at least $T_i$ time units apart (hence the name), which ensures that it causes no more interference than a regular (non-self-suspending) sporadic task.

\subsubsection{Incompatibility with the Dynamic Self-Suspension Model}
\label{sec:incompat}

Before reviewing the classic analysis based on this intuition, we briefly comment on the difficulty of combining period enforcement with the dynamic self-suspension model (Section~\ref{sec:dynamic}).

In short, to be effective, the period enforcer fundamentally requires the segmented self-suspension model~(Section~\ref{sec:segmented})  because it cannot cope with the unpredictable execution times between (the unpredictably many) self-suspensions that jobs may exhibit under the dynamic self-suspension model.

A simple example can explain why the period enforcer algorithm is not compatible with the dynamic self-suspending task model. Consider a trivial system that has only one task with a total execution time $C_1=1$, a total self-suspension length $S_1=1$, and a period and relative deadline of $D_1=T_1=2$. Suppose the first job of task $\tau_1$ arrives at time~$0$, suspends itself for one time unit, and then executes for one time unit. Further suppose the second job of task $\tau_1$ arrives at time $2$, first executes for $0.5$ time units, then suspends for $1$ time unit, and finally executes for $0.5$ time units. With the period enforcer algorithm in place, the second job of task $\tau_1$ starts its execution at time $3$, at which point it will clearly miss its deadline at time $4$.

In this example, the problem is that the eligibility time of the first computation ``segment'' of the second job is determined by the self-suspension pattern of the first job, even though the first job deferred all of its execution, whereas the second job deferred only a part of its execution. Under the more restrictive segmented self-suspension model~(Section~\ref{sec:segmented}), the pattern of self-suspension and computation times is statically fixed; such a mismatch is hence not possible.

Next, we revisit the original analysis of the period enforcer algorithm.

\subsection{Classic Analysis of the Period Enforcer Algorithm}
\label{sec:classic-analysis}

The central notation in Rajkumar's analysis~\cite{Raj:suspension1991} is a deferrable task, which matches our notion of segmented tasks, as already discussed in Section~\ref{sec:segmented}.  Specifically, Rajkumar states that:
\begin{quote}
``With deferred execution, a task $\tau_i$ can execute its $C_i$ units of execution in discrete amounts $C_i^1, C_i^2$, $\ldots$ with suspension in between $C_i^j$ and $C_i^{j+1}$.'' \cite[Section 3]{Raj:suspension1991}\footnote{The notation has been altered here for the sake of consistency. } 
\end{quote}

Central to Rajkumar's analysis~\cite{Raj:suspension1991} is a task set transformation (recall Section~\ref{sec:misc-defs}) that splits each deferrable task with multiple segments (Section~\ref{sec:segmented}) into a corresponding number of single-segment deferrable tasks (Section~\ref{sec:single-segmented}).  In the words of Rajkumar~\cite[Section 3]{Raj:suspension1991}:

\begin{quote}
	 ``Without any loss of generality, we shall assume that a task $\tau_i$ can defer its entire execution time but not parts of it. That is, a task $\tau_i$ executes for $C_i$ units with no suspensions once it begins execution. Any task that does suspend after it executes for a while can be considered to be two or more tasks each with its own worst-case execution time. The only difference is that if a task $\tau_i$ is split into two tasks $\tau_i'$ followed by $\tau_i''$, then $\tau_i''$ has the same deadlines as $\tau_i{{'}}$.''
\end{quote}
In other words, the transformation can be understood as splitting each self-suspending task into a matching number of single-segment deferrable tasks (Section~\ref{sec:single-segmented}), which are equivalent to non-self-suspending sporadic tasks subject to release jitter (Section~\ref{sec:jitter}), which can be easily analyzed with classic fixed-priority response-time analysis~\cite{ABRTW:93}. To constitute an effective schedulability analysis, the transformation must ensure that, if the transformed set of single-segment deferrable tasks can be shown to be schedulable (e.g., with response-time analysis~\cite{ABRTW:93}), then the original set of multi-segment deferrable tasks is also schedulable under period enforcement.

To summarize, as illustrated in Figure \ref{fig:not-ok-without-period-enforcement},
uncontrolled deferred execution can impose  increased interference on lower-priority tasks because of the potential for ``back-to-back'' execution~\cite{LSS:87,LSST:91,Ra:90,ABRTW:93,SLS:95}. The purpose of the period enforcer algorithm is to reduce such penalties for lower-priority tasks without detrimentally affecting the schedulability of self-suspending, higher-priority tasks. The latter aspect --- no detrimental effects for self-suspending tasks --- is captured concisely by Theorem 5 in the original analysis of the period enforcer algorithm \cite{Raj:suspension1991}.
\begin{quote}
{\bf Theorem 5}: A [single-segment] deferrable task that is schedulable under its worst-case conditions is also schedulable under the period enforcer algorithm \cite{Raj:suspension1991}. 
\end{quote}
The ``worst-case conditions'' mentioned in the theorem simply correspond to the case when \textbf{(i)}~a job of a single-segment deferrable task defers its execution for the maximally allowed time $S_i^0$ (i.e., when it incurs maximal release jitter) and \textbf{(ii)} it incurs maximum higher-priority interference (i.e., when its start of execution coincides with a critical instant~\cite{LL:73}).

\subsection{Questions Answered in This Paper}
\label{sec:questions}

Theorem 5 (in \cite{Raj:suspension1991}) is a strong result: it implies that the period enforcer does not induce any deadline misses. This seemingly enables a powerful analysis approach: if the corresponding transformed set of single-segment deferrable tasks can be shown to be schedulable  \emph{without} period enforcement under fixed-priority scheduling using \emph{any} applicable analysis (e.g.,~\cite{ABRTW:93}), then the period enforcer algorithm also yields a correct schedule. 

However, recall that, in the original analysis~\cite{Raj:suspension1991}, deferrable tasks are assumed to defer their  execution either completely or not at all (but not parts of it). It is hence important to realize that Theorem 5 in \cite{Raj:suspension1991} applies only to the transformed set of \emph{single-segment} deferrable tasks, and that it does \emph{not} apply to the \emph{original} set of multi-segmented self-suspending tasks. 

This leads to the first question: \emph{If the  original set of segmented self-suspending tasks is schedulable without period enforcement, is it then also schedulable under period enforcement?} That is, can Theorem 5 (in \cite{Raj:suspension1991}) be generalized to multi-segmented self-suspending tasks? In Section \ref{sec:unschedulable}, we answer this question in the negative.

\begin{enumerate}
	\item There exist sets of segmented self-suspending tasks that are schedulable under fixed-priority scheduling without any enforcement, but that are infeasible under period enforcement. This shows that Theorem 5 in \cite{Raj:suspension1991} has to be  used with care --- it may be applied only in the context of the transformed single-segment deferrable task set, but not in the context of the original multi-segmented self-suspending task set.
\end{enumerate}

Therefore, to apply Theorem 5 to conclude that a set of segmented self-suspending task sets remains schedulable despite period enforcement, we first have to answer the task-set transformation question: \emph{given a set of segmented self-suspending tasks $\tset$, how do we obtain a corresponding set of single-segment deferrable tasks $\tset'$ such that $\tset'$ is schedulable (without period enforcement)  only if $\tset$ is schedulable (with period enforcement)?} That is, as discussed in Section~\ref{sec:classic-analysis}, the classic analysis of the period enforcer \cite{Raj:suspension1991} presumes that it is possible to convert a set of multi-segmented self-suspending tasks into a corresponding set of single-segment deferrable tasks, but it is left undefined in~\cite{Raj:suspension1991}  \emph{how} this central step should be accomplished. In Section \ref{sec:convert}, we make a pertinent observation.

\begin{enumerate}
\setcounter{enumi}{1}
	\item How to derive a single-segment deferrable task set corresponding to a given set of  multi-segmented self-suspending tasks is an open problem. Recent findings by Nelissen et al.~\cite{ecrts15nelissen} can be applied in a special case, but their method takes exponential time (even in the  special case).
\end{enumerate}

Finally, we consider the use of the period enforcer in conjunction with suspension-based multiprocessor locking protocols for partitioned fixed-priority scheduling (such as the MPCP~\cite{LNR:09,Ra:90} or the FMLP~\cite{BLBA:07,BA:08}). While it is certainly tempting to apply period enforcement with the intention of avoiding the negative effects of deferred execution due to lock contention (as previously suggested elsewhere~\cite{Raj:91,Lak:11,LNR:09}), we ask: \emph{does existing blocking analysis remain safe when combined with the period enforcer algorithm?} In Section \ref{sec:locking}, we show that this is not the case.

\begin{enumerate}
\setcounter{enumi}{2}
	\item The period enforcer algorithm invalidates all existing blocking analyses for real-time semaphore protocols as there exist non-trivial feedback cycles between the period enforcer rules and blocking durations.
\end{enumerate}

\section{Period Enforcement Can Induce Deadline Misses}
\label{sec:unschedulable}

In this section, we demonstrate with an example that there exist sets of sporadic  segmented self-suspending tasks that both \textbf{(i)} are schedulable \emph{without} period enforcement and \textbf{(ii)} are not schedulable with period enforcement.

\begin{figure}[t]
  \centering
  \includegraphics[scale=1]{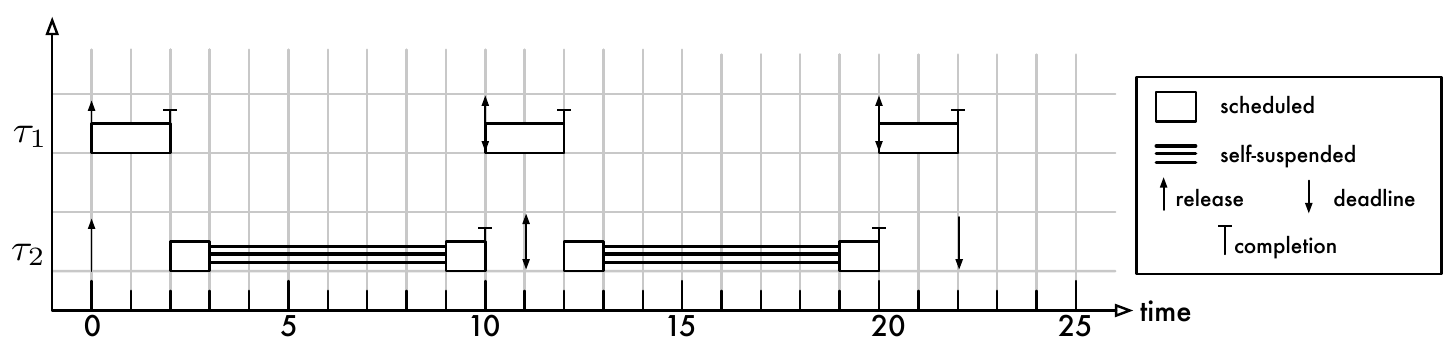}
  \caption{An illustrative example of the original self-suspending task set (without period enforcement) assuming periodic job arrivals on a uniprocessor. Task $\tau_1$ has higher priority than task $\tau_2$.}
  \label{fig:example-original}
  \end{figure}

To this end, consider a task system consisting of $2$ tasks. Let $\tau_1$ denote a sporadic task without self-suspensions and parameters $C_1 = 2$ and $T_1=D_1=10$, and let $\tau_2$ denote a self-suspending task consisting of two segments with parameters  $C_2^1 = 1$,  $S_2^1 = 6$, $C_2^2=1$, and $ T_2=D_2=11$. Suppose that we use the rate-monotonic priority assignment, i.e., $\tau_1$ has higher priority than $\tau_2$. This task set is schedulable without any enforcement since at most one computation segment of a job of $\tau_2$ can be delayed by $\tau_1$: 
\begin{itemize}
	\item if the first segment of a job of $\tau_2$ is interfered with by $\tau_1$, then the second segment resumes at most after $9$ time units after the release of the job and the response time of task $\tau_2$ is hence $10$; otherwise,
	\item  if the first segment of a job of $\tau_2$ is not interfered with by $\tau_1$, then the second segment resumes at most $7$ time units after the release of the job and hence the  response time of task $\tau_2$ is at most $10$ even if the second segment is interfered with by $\tau_1$.
\end{itemize}
Figure \ref{fig:example-original} depicts an example schedule of the task set assuming periodic job arrivals.

\begin{figure}[t]
  \centering
  \includegraphics[scale=1]{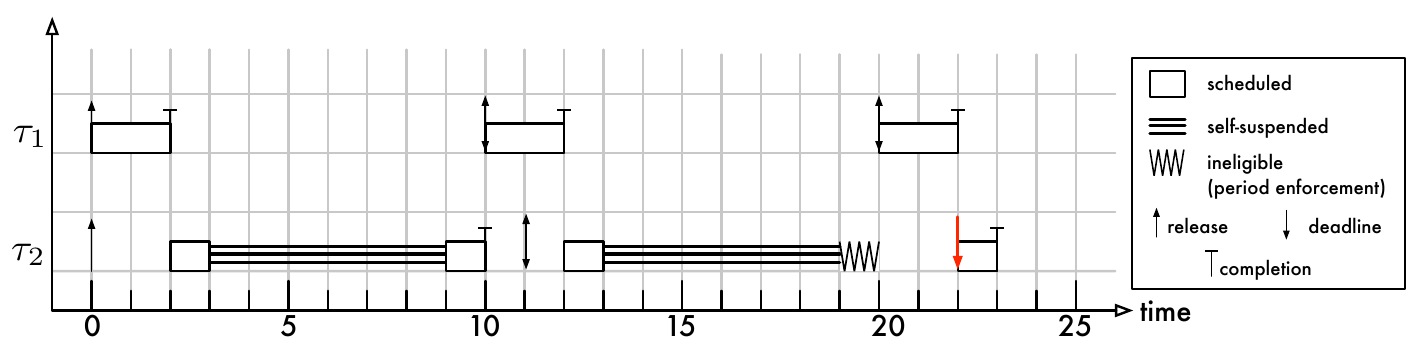}
  \caption{An illustrative example demonstrating a deadline miss at time 22
    under the period enforcer algorithm. At time~19, $\tau_2$ resumes, but it remains ineligible to execute until time~20 when $\tau_1$ is released.}
\label{fig:example}  
\end{figure}

Next, let us consider the same task set under control of the period enforcer algorithm, as defined in Section \ref{sec:pe}.
Figure \ref{fig:example} shows the resulting schedule for a periodic release pattern. The first job of task $\tau_2$ (which arrives at time $a^1_{2,1} =  0$) is executed as if there is no period enforcement since the definition $ET_{2,0}^1 = ET_{2,0}^2 = -T_2$ ensures that both segments are immediately eligible. Note that the first segment of $\tau_2$'s first job is delayed due to interference from $\tau_1$. As a result, the second segment of $\tau_2$'s first job does not resume until time $a^2_{2,1} = 9$. Thus, we have
\begin{align*}
	ET_{2,1}^1 & = \max\left(-T_2 + T_2,\ \mathit{busy}(\tau_2, 0)\right) = 0  \text{ and }
\\
	ET_{2,1}^2 & = \max\left(-T_2 + T_2,\ \mathit{busy}(\tau_2, 9\right) ) = 9.
\end{align*}

In contrast to the first job, the second job of task $\tau_2$ (which is released at time $11$) is affected by period enforcement. The first segment of the second job arrives at time $a^1_{2,2} = 11$, incurs interference for one time unit during $[11, 12)$, and suspends at time $13$. The  second segment of the second job hence resumes only at time $a^2_{2,2} = 19$. Thus, we have
\begin{align*}
	ET_{2,2}^1 & = \max\left(0 + 11,\ \mathit{busy}(\tau_2, 11)\right) = 11  \text{ and }
\\
	ET_{2,2}^2 & = \max\left(9 + 11,\ \mathit{busy}(\tau_2, 19\right) ) = 20.
\end{align*}
According to the rules of the period enforcer algorithm, the processor therefore remains idle at time $19$ because the segment is not eligible to execute until time $ET_{2,2}^2 = 20$. However, at time $20$, the third job of $\tau_1$ is released. As a result, the second job of $\tau_2$ suffers from additional interference and misses its deadline at time $22$.

This example shows that there exist sporadic segmented self-suspending task sets that   \textbf{(i)} are schedulable under fixed-priority scheduling without any enforcement, but \textbf{(ii)} are not schedulable under the period enforcer algorithm.

One may consider to enrich the period enforcer with the following scheduling rule: when the processor becomes idle, a task immediately becomes eligible to execute regardless of its eligibility time. However, even with this extension, the above example remains valid by introducing one additional lower-priority task $\tau_3$ with execution time $C_3=13$ (to be executed from time $3$ to time $9$ and time $13$ to time $20$) and $T_3=D_3=100$. With task $\tau_3$, the processor is always busy from time $0$ to time $23$ and consequently $\tau_2$ still misses its deadline at time $22$.

Furthermore, the example also demonstrates that the conversion to single-segment deferrable tasks does incur a loss of generality since it introduces pessimism. In the context of the above example, if we convert the multi-segmented suspending task $\tau_2$ into two single-segment deferrable tasks, called $\tau_2^1$ and $\tau_2^2$, where task $\tau_2^1$ never defers its execution and task $\tau_2^2$ defers its execution by at most \emph{$9$} time units, the resulting single-segment deferrable task set $\{\tau_1, \tau_2^1, \tau_2^2\}$ is in fact not schedulable under the given priority assignment: if a job of $\tau_1$ coincides with the arrival of a job of $\tau_2^2$ after it has maximally deferred its execution, the job of $\tau_2^2$ has a response time of $9 + 2 + 1$ time units, which exceeds its relative deadline of 11 time units. This shows that any restriction to single-segment deferrable tasks --- that is, assuming that ``[w]ithout any loss of generality [\ldots] a task $\tau_i$ can defer its entire execution time but not parts of it''~\cite{Raj:suspension1991} (recall Section~\ref{sec:classic-analysis}) --- does in fact come with a loss of generality.

\section{Deriving a Corresponding Deferrable Task Set}
\label{sec:convert}

To apply an analysis of the period enforcer based on Theorem 5 in~\cite{Raj:suspension1991}, we first need to convert a given set of multi-segment self-suspending tasks into a corresponding set of single-segment deferrable tasks. This raises the question: how can we efficiently derive the corresponding set of single-segment deferrable tasks? 

The original period enforcer proposal~\cite{Raj:suspension1991} is silent on this issue and does not spell out a procedure for converting a multi-segmented self-suspending task to a corresponding set of single-segment deferrable tasks. However, in our opinion, performing such a transformation without introducing additional pessimism is not at all easy in the general case.

In the following, we illustrate the inherent difficulty of the problem by focusing on a special case to which we can apply a recent result of Nelissen et al. \cite{ecrts15nelissen}, which allows analyzing the exact worst-case response time of multi-segmented self-suspending sporadic tasks, albeit with exponential time complexity. 
Nelissen et al.'s worst-case response time analysis~\cite{ecrts15nelissen} is exact under the following conditions:\footnote{We refer to the characteristics of the worst-case release pattern provided in Lemma 2 in \cite{ecrts15nelissen}. The exact worst-case response time can be obtained by exploring all  release patterns that satisfy these conditions.} 
\begin{enumerate}
	\item the task set contains only one self-suspending task, 
	\item the self-suspending task is the lowest-priority task, 
	\item the scheduling policy is preemptive fixed-priority scheduling, and 
	\item all tasks have constrained deadlines (i.e., $D_i \leq T_i$ for all $\tau_i$).
\end{enumerate}

For an arbitrary number of tasks $k \geq 2$, 
suppose that the system has $k-1$ regular sporadic tasks and only one segmented self-suspending task $\tau_k$, and that all tasks have implicit deadlines (i.e., $D_i = T_i$ for all $\tau_i$). Further suppose that task $\tau_k$ has $m_k$ segments with $m_k \geq 3$.  

To  convert a computation segment of $\tau_k$ into a single-segment deferrable task, we need to derive the segment's \emph{latest-possible arrival time}, relative to the release of a job. Formally,  for the $j^{\mathrm{th}}$ computation segment of task $\tau_k$, we let $\rho_k^j$ denote its latest-possible arrival time, with the interpretation that, if a job of task $\tau_k$ arrives at time $t$, then  it is guaranteed that the $j^{\mathrm{th}}$ computation segment of this job will not arrive later than at time $t+\rho_k^j$.

How can we compute $\rho_k^j$? Suppose that the worst-case response time of the $j^{\mathrm{th}}$ computation segment of task $\tau_k$ is $W_k^j$, and recall that $S_k^{j}$ denotes the maximum self-suspension length before the $j^{\mathrm{th}}$ computation segment of $\tau_k$. Then $\rho_k^j$ can be expressed in terms of $W_k^{j-1}$:
$$
	\rho_k^j = W_k^{j-1}+S_k^{j-1},
$$
where $W_k^0 = 0$.  Therefore, if we can derive the exact segment worst-case response time $W_k^j$ for $j=1,2,\ldots,m_k-1$, we can easily compute $\rho_k^j$  for $j=1,2,\ldots,m_k$. And conversely, if we can somehow obtain $\rho_k^j$  for $j=2,\ldots,m_k$, we  can trivially infer $W_k^j$ for $j=1,2,\ldots,m_k-1$.
Based on these considerations, it appears that the transformation problem is  --- at least in the considered special case --- equivalent to the  worst-case response time analysis of a multi-segmented self-suspending task. 

However, deriving an exact bound $W_k^j$ for $j=1,2,\ldots,m_k-1$ for task $\tau_k$ is not easy: 
even for the above ``simple'' case, Nelissen et al.'s solution~\cite{ecrts15nelissen} for calculating the exact worst-case response time requires exponential time complexity if $j \geq 2$. Furthermore, Nelissen et al. \cite{ecrts15nelissen} identified several misconceptions in prior analyses, and after correcting those misconceptions, observed that the problem of deriving the worst-case response time of a computation segment in pseudo-polynomial time seems to be very challenging indeed.\footnote{In fact, in ongoing work, it has recently been shown that verifying the schedulability of task $\tau_k$ is coNP-hard in the strong sense even in the considered simplified case~\cite{Chen2016b}.}

Nelissen et al.~\cite{ecrts15nelissen} did not study the period enforcer; rather, they considered unrestricted self-suspensions. However, given that the period enforcer has no effect on tasks that do not self-suspend~\cite{Raj:suspension1991}, and given that in the considered special case only the lowest-priority task self-suspends, we believe that these observations transfer to the period enforcement case.

To summarize, to analyze the period enforcer based on Theorem~5 in~\cite{Raj:suspension1991}, a procedure for transforming multi-segmented self-suspending tasks into sets of single-segment deferrable tasks is needed, but no such procedure is given in the original proposal~\cite{Raj:suspension1991}.
Based on the presented considerations, we conclude that filling in this missing step is non-trivial and observe that the closest known solution by Nelissen et al.~\cite{ecrts15nelissen} requires exponential time even in the greatly simplified special case of a single self-suspending task. It thus remains unclear how Theorem~5 in~\cite{Raj:suspension1991} can be used for schedulability analysis of sets of multi-segmented self-suspending tasks. 
While we did search for alternative analysis approaches that do not rely on Theorem 5, we did not find a simple or efficient schedulability test for the period enforcer without introducing substantial additional pessimism.  The problem remains open. 

Next, we take a look at the period enforcer in the context of synchronization protocols.

 \section{Incompatibility with Suspension-Based Locking Protocols}
\label{sec:locking}

Binary semaphores, i.e., suspension-based locks used to realize mutually exclusive access to shared resources, are a common source of self-suspensions in multiprocessor real-time systems. When a task tries to use a resource that has already been locked, it self-suspends until the resource becomes available. Such self-suspensions due to lock contention, just like any other self-suspension, result in deferred execution and thus can detrimentally affect a task's interference on lower-priority tasks. It may thus seem natural to apply the period enforcer  to control the negative effects of blocking-induced self-suspensions.\footnote{The use of  period enforcement in combination with suspension-based locks has indeed been assumed in prior work~\cite{Raj:91}, stated as a motivation and possible use case in the original period enforcer proposal~\cite{Raj:suspension1991}, and suggested as a potential improvement elsewhere~\cite{Lak:11,LNR:09}.} However, as we demonstrate with two examples, it is actually unsafe to apply period enforcement to lock-induced self-suspensions.

\subsection{Combining Period Enforcement and Suspension-Based Locks}

Whenever a task attempts to lock a shared resource, it may potentially block and self-suspend. In the context of the multi-segmented self-suspending task model, each lock request hence marks the beginning of a new segment.

The period enforcer algorithm may therefore be applied to determine the eligibility time of each such segment (which, again, all start with a critical section). There is, however, one complication: when does a task actually \emph{acquire} a lock? That is, if a task's execution is postponed due to the period enforcement rules, at which point is the lock request processed, with the consequence that the resource becomes unavailable to other tasks? 

There are two possible interpretations of how period enforcement and locking rules may interact. Under the \textbf{first interpretation}, when a task requires a shared resource, which implies the beginning of a new segment, its lock request is processed \emph{only when its new segment is eligible for execution}, as determined by the period enforcer algorithm. Alternatively, under the \textbf{second interpretation}, a task's request is processed \emph{immediately} when it requires a shared resource.

As a consequence of the first rule, a task may find a required shared resource unavailable when its new segment becomes eligible for execution even though the resource was available when the prior segment finished.  As a consequence of the second rule, a shared resource may be locked by a task that cannot currently use the resource  because the task is still ineligible to execute.

We believe that the first interpretation is the more natural one, as it does not make much sense to allocate resources to tasks that cannot yet use them. However, for the sake of completeness, we show that either interpretation can lead to deadline misses even if the task set is trivially schedulable without any enforcement.

\subsection{Case 1: Locking Takes Effect at Earliest Segment Eligibility Time}
\label{sec:sem-case-1}
In the following example, we assume the first interpretation, i.e., that the processing of lock requests is delayed until the point when a resuming segment would no longer be subject to any delay due to period enforcement. We show that this interpretation leads to a deadline miss in a task set that would otherwise be trivially schedulable.

Consider the following simple task set consisting of two tasks on two processors that share one resource. Task $\tau_1$, on processor~1, has a total execution cost of $C_1 = 4$ and a period and deadline of $T_1 = D_1 = 8$. After one time unit of execution, jobs of $\tau_1$ require the shared resource for two time units. $\tau_1$ thus consists of two segments with costs $C_1^1 = 1$ and $C_1^2 = 3$. Task $\tau_2$, on processor~2, has the same overall WCET ($C_2 = 4$), a slightly shorter period ($T_2 = D_2 = 7$), and requires the shared resource for one time unit after \emph{two} time units of execution ($C_2^1 = 2$ and $C_2^2 = 2$). Without period enforcement (and under any reasonable locking protocol), the task set is trivially schedulable because, by construction, any job of $\tau_1$ incurs at most one time unit of blocking, and any job of $\tau_2$ incurs at most two time units of blocking.

\begin{figure}[t]
  \centering
  \includegraphics[scale=1]{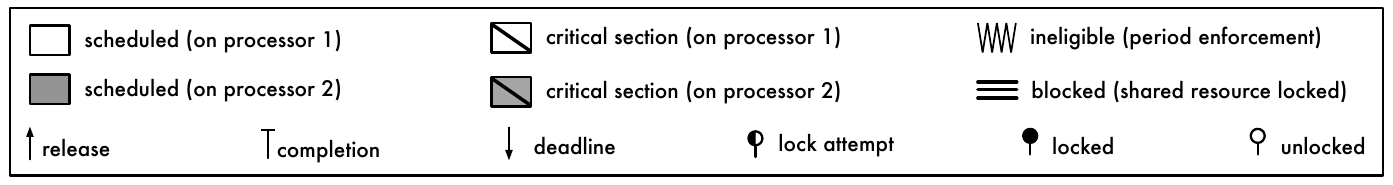}
  \includegraphics[scale=1]{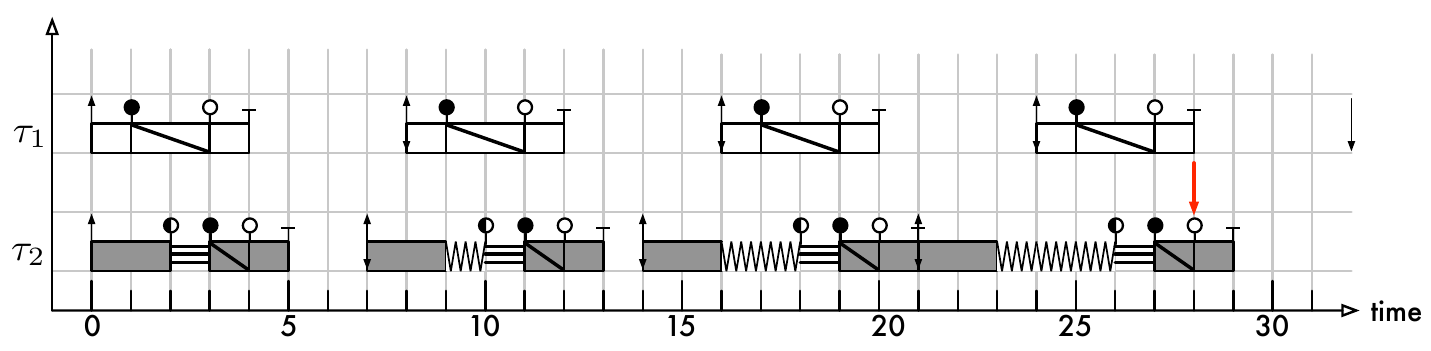}
  \caption{Example schedule of two tasks $\tau_1$ and $\tau_2$ on two processors sharing one lock-protected resource. The example assumes that lock requests take effect only when the critical section segment  becomes eligible to be scheduled according to the rules of the period enforcer algorithm. Under this interpretation, the fourth job of task $\tau_2$ misses its deadline at time $28$.}
  \label{fig:locking-alt1}
\end{figure}

In contrast, with period enforcement, deadline misses are possible.
Figure~\ref{fig:locking-alt1} depicts a schedule of the two tasks assuming periodic job arrivals and use of the period enforcer algorithm. We focus on the eligibility times $ET_{2,1}^2,ET_{2,2}^2,ET_{2,3}^2,\ldots$ of the second segment of $\tau_2$.

Since $\tau_2$'s first job requests the shared resource only after two time units of execution, it is blocked by $\tau_1$'s critical section, which commenced at time $1$. At time $3$, $\tau_1$ releases the shared resource and $\tau_2$ consequently resumes (i.e., $a^2_{2,1} = 3$). According to the period enforcer rules~\cite{Raj:suspension1991}, the second segment is immediately eligible because, according to Equation~\ref{eq:ET-def} (in Section~\ref{sec:unschedulable}),
\begin{align*}
	ET_{2,1}^2 & = \max\left(ET_{2,0}^2 + T_2,\ \mathit{busy}(\tau_2, a^2_{2,1})\right) =\max(-T_2 + T_2,\ 3) = 3.
\end{align*}
(Recall that $ET_{2,0}^2 = -T_2$, and interpret $\mathit{busy}(\tau_2, a^2_{2,1})$ with respect to $\tau_2$'s processor.)

At time $7$, the second job of $\tau_2$ is released. Its first segment ends at time $9$. However, its second segment is not eligible to be scheduled before time $10$ since $ET_{2,2}^2 \geq ET_{2,1}^2 + T_2 = 3 + 7 = 10$. At time $9$, the second job of $\tau_1$, released at time $8$, can thus lock the shared resource without contention. Consequently, when $\tau_2$'s request for the shared resource takes effect at time $10$, the resource is no longer available and $\tau_2$ must wait until time $a^2_{2,2} = 11$ before it can proceed to execute. We thus have
\begin{align*}
	ET_{2,2}^2 & = \max\left(ET_{2,1}^2 + T_2,\ \mathit{busy}(\tau_2, a^2_{2,2})\right) =\max(10,\ 11) = 11.
\end{align*}

The third job of $\tau_2$ is released at time $14$. Its first segment ends at time $16$, but since $ET_{2,3}^2 \geq ET_{2,2}^2 + T_2 = 11 + 7 = 18$, the second segment may not commence execution until time~$18$ and the shared resource remains available to other tasks in the meantime. The third job of $\tau_1$ is released at time $16$ and acquires the uncontested shared resource at time $17$. Thus, the segment of $\tau_2$ cannot resume execution before time $a^2_{2,3} = 19$. Therefore
\begin{align*}
	ET_{2,3}^2 & = \max\left(ET_{2,2}^2 + T_2,\ \mathit{busy}(\tau_2, a^2_{2,3})\right) =\max(18,\ 19) = 19.
\end{align*}

The same pattern repeats for the fourth job of $\tau_2$, released at time $21$: when its first segment ends at time $23$, the second segment is not eligible to commence execution before time $26$ since $ET_{2,4}^2 \geq ET_{2,3}^2 + T_2 = 19 + 7 = 26$. By then, however, $\tau_1$ has already locked the shared semaphore again, and the second segment of the fourth job of $\tau_2$ cannot resume before time $a^2_{2,4} = 27$, at which point
\begin{align*}
	ET_{2,4}^2 & = \max\left(ET_{2,3}^2 + T_2,\ \mathit{busy}(\tau_2, a^2_{2,4})\right) =\max(26,\ 27) = 27.
\end{align*}
However, this leaves insufficient time to meet the job's deadline: as the second segment of $\tau_2$ requires $C_2^2 = 2$ time units to complete, the job's deadline at time~$28$ is  missed.

By construction, this example does not depend on a specific locking protocol; for instance, the effect occurs with both the MPCP~\cite{Ra:90} (based on priority queues) and the FMLP~\cite{BLBA:07,BA:08} (based on FIFO queues).  The corresponding response-time analyses for both protocols~\cite{Br:13,LNR:09} predict a worst-case response time of $6$ for task $\tau_2$ (i.e., four time units of execution, and at most two time units of blocking due to the critical section of $\tau_1$). 
This demonstrates that, under the first interpretation, adding period enforcement to suspension-based locks invalidates existing blocking analyses. Furthermore, it is clear that the devised repeating pattern can be used to construct schedules in which the response time of $\tau_2$  grows beyond any given implicit or constrained deadline.

Next, we show that the second interpretation can also lead to deadline misses in otherwise trivially schedulable task sets.

\subsection{Case 2: Locking Takes Effect Immediately}
\label{sec:sem-case-2}
 From now on, we assume the second interpretation: all lock requests are processed immediately when they are made, even if this causes the shared resource to be locked by a task that is not yet eligible to execute according to  the rules of the period enforcer algorithm. We construct an example in which a task's response time grows with each job until a deadline is missed.

To this end, consider two tasks with identical parameters hosted on two processors. Task $\tau_1$ is hosted on processor~1; task $\tau_2$ is hosted on processor~2. Both tasks have the same period and relative deadline $T_1 = T_2 = D_1 = D_2 = 8$ and the same WCET of $C_1 = C_2 = 4$. They both access a single shared resource for two time units each per job. Both tasks request the shared resource after executing for \emph{at most} one time unit. They both thus have two segments each with parameters $C_1^1 = C_2^1 = 1$ and $C_1^2 = C_2^2 = 3$. 

\begin{figure}[t]
  \centering
  \includegraphics[scale=1]{legend.pdf}
  \includegraphics[scale=1]{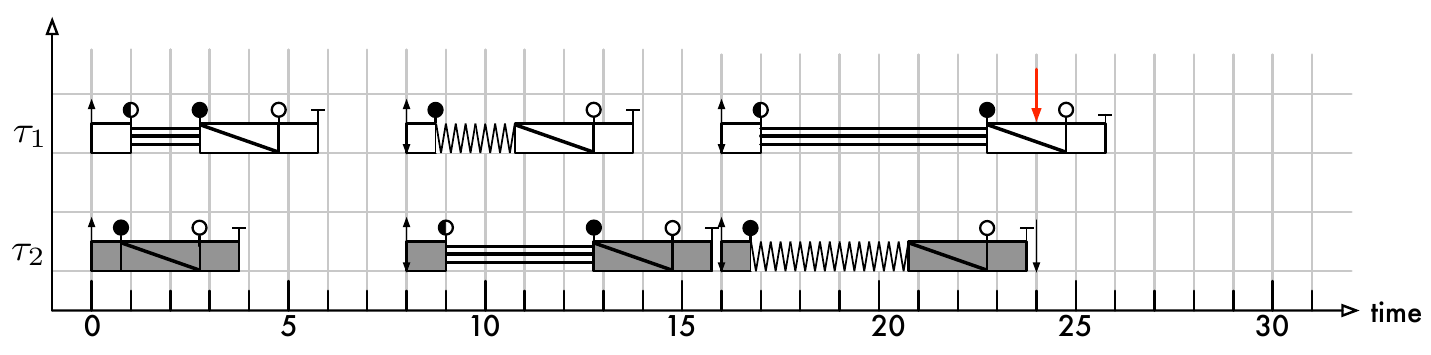}
  \caption{Example schedule of two tasks $\tau_1$ and $\tau_2$ on two processors sharing one lock-protected resource. The example assumes that lock requests take effect immediately, even if the critical section segment is not yet eligible to be scheduled according to the rules of the period enforcer algorithm. Under this interpretation, the third job of task $\tau_1$ misses its deadline at time $24$.}
  \label{fig:locking-alt2}
\end{figure}

The example exploits that a job may require \emph{less} service than its task's specified WCET. To ensure that the shared resource is acquired in a certain order, we assume the following deterministic pattern of the actual execution times. Let $\epsilon$ be an arbitrarily small, positive real number with $\epsilon <1$. 
\begin{itemize}
	\item The first segment of even-numbered jobs  of $\tau_1$ executes for only $1-\epsilon$ time units.
	\item The first segment of odd-numbered jobs of $\tau_2$ executes for only $1-\epsilon$ time units.
	\item All other segments execute for their specified worst-case costs.
\end{itemize}
Figure~\ref{fig:locking-alt2} shows an example schedule assuming periodic job arrivals.

At time $1-\epsilon$, the first job of $\tau_2$ acquires the shared resource because $\tau_1$ does not issue its request until time $1$. Consequently, $\tau_1$ is blocked until time $a^2_{1,1} = 3 - \epsilon$, and we have
\begin{align*}
	ET_{1,1}^2 & = \max\left(ET_{1,0}^2 + T_1,\ \mathit{busy}(\tau_1, a^2_{1,1})\right) =\max(-T_1 + T_1,\ 3 - \epsilon) = 3 - \epsilon
\\ \intertext{and}
	ET_{2,1}^2 & = \max\left(ET_{2,0}^2 + T_2,\ \mathit{busy}(\tau_2, a^2_{2,1})\right) =\max(-T_2 + T_2,\ 0) = 0.
\end{align*}

The roles of the second jobs of both tasks are reversed: since the second job of $\tau_1$ locks the shared resource already at time $9-\epsilon$, $\tau_2$ is blocked when it attempts to lock the resource at time~$9$. However, according to the rules of the period enforcer algorithm, the second segment of the second job of $\tau_1$ is not actually eligible to execute before time $11 - \epsilon$ since
\begin{align*}
	ET_{1,2}^2 & = \max\left(ET_{1,1}^2 + T_1,\ \mathit{busy}(\tau_1, a^2_{1,2})\right) =\max(3 - \epsilon + 8,\ 8) = 11 - \epsilon.
\end{align*}
Consequently, even though the lock is granted to $\tau_1$ already  at time $9-\epsilon$, the critical section is executed only starting at time $11 - \epsilon$, and $\tau_2$ is thus delayed until time $13 - \epsilon$. At time $13 - \epsilon$, $\tau_2$ is immediately eligible to execute since
\begin{align*}
	ET_{2,2}^2 & = \max\left(ET_{2,1}^2 + T_2,\ \mathit{busy}(\tau_2, a^2_{2,2})\right) =\max(0 + 8,\ 13 - \epsilon) = 13 - \epsilon.
\end{align*}

The third jobs of both tasks are released at time $16$. The roles are swapped again: because $\tau_2$'s first segment requires only $1-\epsilon$ time units of service, it acquires the lock at time $a^2_{2,3} = 17 - \epsilon$, before $\tau_1$ issues its request at time~$17$. However, according to the period enforcer algorithm's eligibility criterium, $\tau_2$ cannot actually continue its execution before time $21- \epsilon$ since
\begin{align*}
	ET_{2,3}^2 & = \max\left(ET_{2,2}^2 + T_2,\ \mathit{busy}(\tau_2, a^2_{2,3})\right) =\max(13- \epsilon + 8,\ 16) = 21- \epsilon.
\end{align*}
This, however, means that $\tau_1$ cannot use the shared resource before time $23 - \epsilon$, which leaves insufficient time to complete the second segment of $\tau_1$'s third job before its deadline at time $24$.
Furthermore, if both tasks continue the illustrated execution pattern, the period enforcer continues to increase their response times. As a result, the pattern may be repeated to construct schedules in which any arbitrarily large implicit or constrained deadline is violated.

As in the previous example,  the response-time analyses for both the MPCP~\cite{Br:13,LNR:09} and the   FMLP~\cite{Br:13} predict a worst-case response time of $6$ for both tasks (i.e., four time units of execution, and at most two time units of blocking). The example thus demonstrates that, if lock requests take effect immediately, then the period enforcer is incompatible with existing blocking analyses because, under the second interpretation, it increases the effective lock-holding times.

\subsection{Other Protocols and Interpretations}

The examples in Sections~\ref{sec:sem-case-1} and~\ref{sec:sem-case-2} assume a \emph{shared-memory} locking protocol: once a lock is granted, tasks execute their own critical sections on their assigned processors. One may wonder whether  effects similar to those described in Sections~\ref{sec:sem-case-1} and~\ref{sec:sem-case-2} can also occur under \emph{distributed} real-time locking protocols such as the \emph{Distributed Priority Ceiling Protocol} (DPCP)~\cite{RSL:88,Raj:91} or the \emph{Distributed FIFO Locking Protocol} (DFLP)~\cite{Br:13,Br:14}, where critical sections may be executed on dedicated \emph{synchronization processors}. In this case, the self-suspension occurs on the task's \emph{application processor}, which is different from the (remote) synchronization processor on which the critical section is executed.

This separation allows employing period enforcement  only on application processors (while avoiding it on synchronization processors) without incurring the feedback cycle between blocking times and self-suspension times highlighted in Sections~\ref{sec:sem-case-1} and~\ref{sec:sem-case-2}.

\begin{figure}[t]
  \centering
  \includegraphics[scale=1]{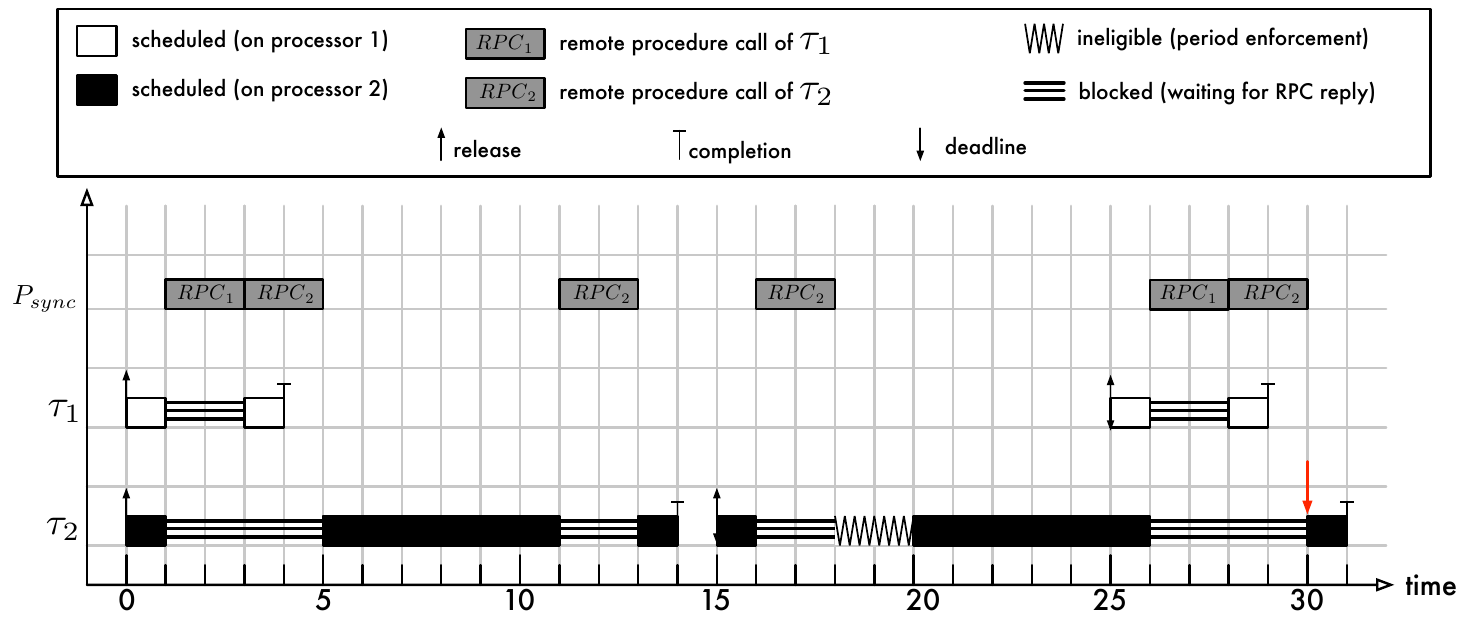}
  \caption{Example schedule of two tasks $\tau_1$ and $\tau_2$ on two processors sharing a \emph{remote} resource using a distributed semaphore protocol (e.g., the DPCP~\cite{RSL:88,Raj:91} or the DFLP~\cite{Br:13,Br:14}) together with the period enforcer. Since critical sections are executed as \emph{remote procedure calls} (RPCs) on a dedicated synchronization processor $P_{sync}$ (not subject to period enforcement), their execution is not delayed by period enforcement. However, period enforcement delays $\tau_2$ at time~18, which invalidates existing analyses~\cite{RSL:88,Raj:91,Br:13}: under both the DPCP and the DFLP, $\tau_2$ is predicted to have a maximum response time of~14~\cite{Br:13}, but with period enforcement, the second job of $\tau_2$ has in fact a response time of 16.
  }
  \label{fig:locking-dpcp}
\end{figure}

However, period enforcement still invalidates all existing blocking analyses for distributed real-time semaphore protocols~\cite{RSL:88,Raj:91,Br:13} because it artificially increases blocking times if tasks contain multiple accesses to shared resources. An example demonstrating this effect is shown in Figure~\ref{fig:locking-dpcp}. Two segmented self-suspending tasks $\tau_1$ and $\tau_2$ share a resource using a distributed real-time locking protocol. The choice of protocol is irrelevant; the example works with both the DPCP and the DFLP. The tasks have parameters $m_1 = 2$, $C_1^1 = C_1^2 = 1$, $T_1 = D_1 = 25$ and $m_2 = 3$, $C_2^1 = C_2^3 = 1$, $C^2_2 = 6$, and $T_2 = D_2 = 15$. The computation segments are separated by self-suspensions that arise while the tasks wait for the completion of critical sections that are executed remotely on a dedicated synchronization processor $P_{sync}$; the corresponding suspension segment parameters $S_1^1$, $S_2^1$, and $S_2^2$ will be defined shortly.

The first jobs of $\tau_1$ and $\tau_2$ are both released at time~0 and attempt to access the shared resource at time~1. Task $\tau_1$'s request is serviced first; as a result $\tau_2$ resumes only at time $a^2_{2,1} = ET_{2,1}^2 = 5$ after having been suspended for four time units:
\begin{align*}
	ET_{2,1}^2 & = \max\left(ET_{2,0}^2 + T_2,\ \mathit{busy}(\tau_2, a^2_{2,1})\right) =\max(-T_2 + T_2,\ 5) = 5.
\end{align*}
Task $\tau_2$ then executes its second computation segment for $C_2^2 = 6$ time units until time~11, when the job accesses the shared resource for a second time. Since there is no contention from $\tau_1$ at this time, $\tau_2$ resumes after only two time units at time~13. This leaves the job sufficient time to complete at time~14, one time unit before its deadline at time~15.

The second job of $\tau_2$ is released at time~15 and issues a request for the shared resource at time~16. Since there is no contention from $\tau_1$ at the time, the second computation segment arrives already at time $a^2_{2,2} = 18$, after having been self-suspended for only two time units. However, since the second segment of the first job arrived at time $a^2_{2,1} = 5$, the second segment of the second job is not eligible to start execution until time $ET_{2,2}^2 = 20$ since 
\begin{align*}
	ET_{2,2}^2 & = \max\left(ET_{2,1}^2 + T_2,\ \mathit{busy}(\tau_2, a^2_{2,2})\right) =\max(5 + T_2,\ 18) = 20.
\end{align*}
As a result, $\tau_2$ faces contention from $\tau_1$ when it issues its second request for the shared resource at time~26, which ultimately leads to a deadline miss at time~30.

In contrast, without period enforcement, $\tau_2$ does not miss its deadline at time~30 because, \emph{across its two requests}, a job of $\tau_2$ is delayed by at most one request of $\tau_1$, for a total self-suspension time of at most two six time units. That is, even though the \emph{individual} self-suspension segments of the two tasks are each up to four time units long (i.e., $S_1^1 = S_2^1 = S_2^2 = 4$), the fact that self-suspensions arise due to the same cause (resource contention) means that the \emph{total} self-suspension time is actually less than the sum of the individual per-segment bounds.

 Existing analyses for the DPCP~\cite{RSL:88,Raj:91,Br:13} and the DFLP~\cite{Br:13} exploit this knowledge and therefore predict task $\tau_2$ to be schedulable with a worst-case response time of 14.\footnote{The analyses in \cite{RSL:88,Raj:91} do assume a segmented task model, but bound the total blocking across all segments. The analysis in \cite{Br:13} also bounds the total blocking across all segments and can be applied to both the segmented and the dynamic self-suspension model.} The example in Figure~\ref{fig:locking-dpcp} thus demonstrates that the period enforcer invalidates existing blocking bounds for distributed semaphore protocols. As an aside, the example in Figure~\ref{fig:locking-dpcp} further highlights limitations of the segmented self-suspension model in the context of synchronization protocols, where the lengths of self-suspensions encountered at runtime are inherently not independent.

Returning to the shared-memory case,
as a third possible interpretation, one could also exclude critical sections from period enforcement such that only the rest of the computation segment \emph{after} a critical section is subject to period enforcement (i.e., making critical sections immediately eligible to execute).\footnote{This interpretation does not fit the assumptions stated in~\cite{Raj:suspension1991,Raj:91}.} This can be understood as making each critical section an individual computation segment (exempt from period enforcement) that is separated from the following computation by a ``virtual'' self-suspension of maximum length zero. As in the case of distributed semaphore protocols, this interpretation breaks the feedback cycle highlighted in Sections~\ref{sec:sem-case-1} and~\ref{sec:sem-case-2}, but still invalidates all existing blocking analyses as it artificially inflates the synchronization delay.

\begin{figure}[t]
  \centering
  \includegraphics[scale=1]{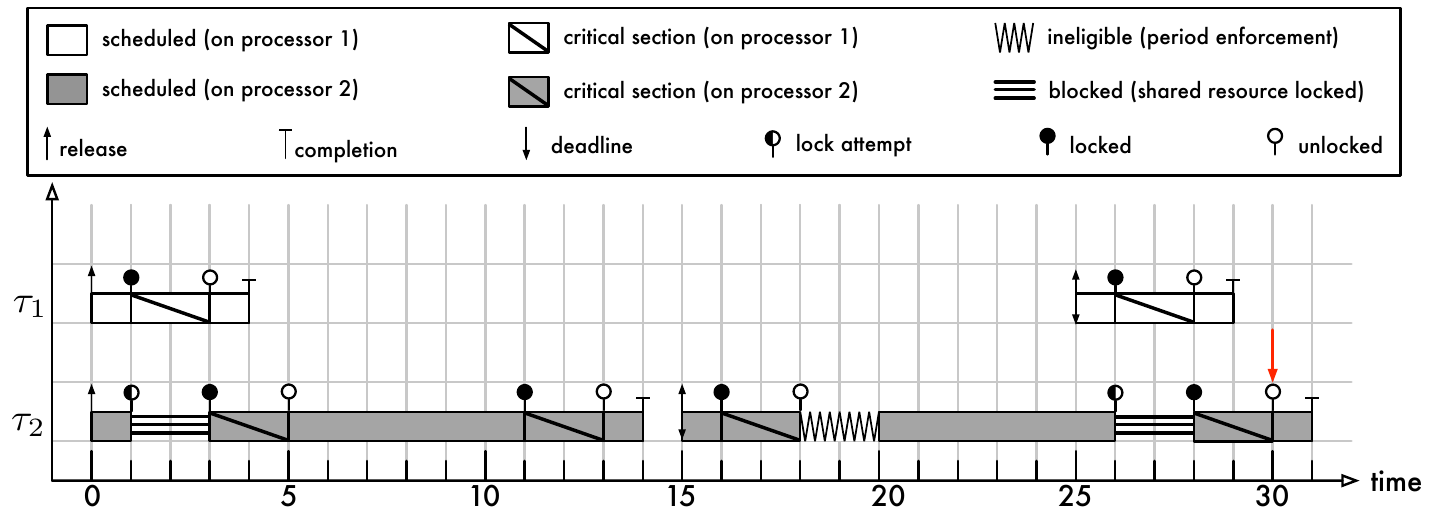}
  \caption{Example schedule of two tasks $\tau_1$ and $\tau_2$ on two processors sharing one lock-protected resource. The example assumes that lock requests take effect immediately and that critical sections are exempt from period enforcement (i.e., period enforcement is applied only to any computation after a critical section). As in Figure~\ref{fig:locking-dpcp}, the second job of task $\tau_2$ misses its deadline at time $30$.
  }
  \label{fig:locking-alt3}
\end{figure}

An example of this effect is shown in  Figure~\ref{fig:locking-alt3}, which depicts the same scenario as in Figure~\ref{fig:locking-dpcp} under the assumption that a shared-memory semaphore protocol is used (i.e., critical sections are executed locally by each job) and that critical sections are exempt from period enforcement. As in the distributed case, period enforcement induces a deadline miss, whereas existing blocking analyses~\cite{Br:13,LNR:09} exploit the fact that a remote critical section can block only once, thus arriving at a worst-case response time bound of 14 for $\tau_2$.

To conclude, in both Figures~\ref{fig:locking-dpcp} and~\ref{fig:locking-alt3}, period enforcement has an effect  \emph{as if} a single remote critical section can block a given job multiple times, which is fundamentally incompatible with efficient blocking analysis~\cite{Br:13}.

\subsection{Discussion}

While it is intuitively appealing to combine period enforcement with suspension-based locking protocols~\cite{Raj:91,Lak:11,LNR:09}, we observe that this causes non-trivial difficulties. In particular, our examples show that the addition of period enforcement invalidates all existing blocking analyses.

If critical sections are subject to period enforcement, our examples also suggest that devising a correct blocking analysis would be a substantial challenge due to the demonstrated feedback cycle between the period enforcer rules and blocking durations. Fundamentally, the design of the period enforcer algorithm implicitly rests on the assumption that a segment \emph{can} execute as soon as it is eligible to do so. In the presence of locks, however, this assumption is invalidated. As demonstrated, the result can be a successive growth of self-suspension times that proceeds until a deadline is missed.  The period enforcer algorithm, at least as defined and used in the literature to date~\cite{Raj:suspension1991,Raj:91}, is therefore incompatible with the existing literature on suspension-based real-time locking protocols (e.g., \cite{Raj:91,Lak:11,LNR:09,BLBA:07,Br:13}). 

Finally, it is worth noting that our examples can be trivially extended with lower-priority tasks to ensure that no processor idles before the described deadline misses occur. It is also not difficult to extend the examples in Figures~\ref{fig:locking-alt2} and~\ref{fig:locking-alt3} with a task on a third processor such that all critical sections of $\tau_1$ and $\tau_2$ are separated from their predecessor segments by a non-zero self-suspension.
 
\section{Concluding Remarks}
\label{sec:conclusion}

We have revisited the underlying assumptions and limitations of the period enforcer algorithm, which Rajkumar \cite{Raj:suspension1991} introduced to handle segmented self-suspending real-time tasks. 

One key assumption in the original proposal \cite{Raj:suspension1991} is that a deferrable task $\tau_i$ can defer its entire execution time but not parts of it. This creates some mismatches between the original segmented self-suspending task set and the corresponding single-segment deferrable task set, which we have demonstrated with an example that shows that Theorem 5 in \cite{Raj:suspension1991} does not reflect the schedulability of the original segmented self-suspending task system.

The original proposal \cite{Raj:suspension1991} further left open the question of how to convert a segmented self-suspending task set to a corresponding set of single-segment deferrable tasks. This problem remains open. Taking into account recent developments~\cite{Chen2016b,ecrts15nelissen}, we have observed that such a transformation is non-trivial in the general case.  

Finally, we have demonstrated that substantial difficulties arise if one attempts to combine suspension-based locks with period enforcement. These difficulties stem from the fact that period enforcement can increase contention or lock-holding times, which increases the lengths of self-suspension intervals, which then in turn feeds back into the period enforcer's minimum suspension lengths. As a consequence, period enforcement invalidates all existing blocking analyses.

Nevertheless, the period enforcer algorithm \emph{per se}, and Theorem 5 in \cite{Raj:suspension1991},  could still prove to be useful for handling self-suspending tasks (that do not use suspension-based locks) if  \emph{efficient} schedulability tests or methods for constructing sets of single-segment deferrable tasks can be found. However, such tests or transformations have not yet been obtained  and the development of a precise and efficient schedulability test for self-suspending tasks remains an open problem.

\section*{Acknowledgements}

We thank James H.\ Anderson and Raj Rajkumar for their comments on early drafts of this paper.
This work has been supported by DFG, as part of the Collaborative
Research Center SFB876. 
 
\bibliography{biblio}{}

\end{document}